\documentclass[prb,superscriptaddress,floatfix,twocolumn,showpacs,amsfonts,amsmath,amssymb]{revtex4}
\usepackage{graphicx} 
\usepackage{dcolumn} 
\usepackage{bm} 
\usepackage{color}
\usepackage{float}

\begin{document}
	
\title{Dimensional reduction of the Luttinger Hamiltonian and g-factors
of holes in symmetric two-dimensional semiconductor heterostructures}

	\date{\today}
	
	\author{D.~S.~Miserev, O.~P.~Sushkov} 
	\affiliation{School of Physics, University of New South Wales, Sydney, Australia}
	
	\begin{abstract}		
The spin-orbit interaction of holes in zinc-blende  semiconductors is much 
stronger than that of electrons. This makes the hole systems very
attractive for possible spintronics applications.
In three dimensions (3D) dynamics of holes is described by 
well known Luttinger Hamiltonian. However, most of recent spintronics 
applications are related to two dimensional heterostructures where dynamics 
in one
direction is frozen due to quantum confinement. The confinement results
in dimensional reduction of the Luttinger Hamiltonian, 3D$\to$2D.
Due to interplay of the spin-orbit interaction,
the external magnetic field, and the lateral gate potential imposed on
the heterostructure the reduction is highly nontrivial and not
known.
In the present work we perform the reduction  and hence derive the 
general effective Hamiltonian which describes  spintronics effects
in symmetric two-dimensional (2D) heterostructures.
In particular, we do the following,
(i) derive the spin-orbit interaction and the Darwin interaction
related to the
lateral gate potential, (ii) determine the momentum dependent 
out-of-plane g-factor, (iii) point out that there are two independent 
in-plane g-factors, (iv) determine momentum dependencies of the in-plane 
g-factors.
	\end{abstract}
	
	\maketitle

\section{Introduction}
Spin orbit interaction (SOI) in cubic zinc-blende  semiconductors
 is of topical interest because
of various spintronics applications and devises.
It is well understood that SO interaction is very different for the conduction
band (electrons) and for the valence band (holes).
Electrons in the conduction band originate from atomic s-orbitals and
therefore they have spin 1/2. 
There are two SOI for electrons, the Dresselhaus interaction~\cite{dresselhaus}
which is due to inversion asymmetry in bulk and Rashba interaction~\cite{Rashba}
which is usually due to asymmetric 2D interfaces.
The SO interactions are relatively week in  semiconductors
with light atoms like GaAs, and they can become significant in  
semiconductors with heavy atoms like InAs and InSb~\cite{kallaher,heida}.

The valence band (holes) originates from atomic p$_{3/2}$ orbitals,
therefore the effective spin is $S=3/2$, and hence the SOI quadratic
in spin is possible. This SOI is always strong,
even in Si it is comparable with kinetic energy. In this sense holes
in zinc-blende  semiconductors are always 'ultrarelativistic',
the SOI is comparable or even larger than kinetic energy.
The quadratic in spin SOI is described by Luttinger 
Hamiltonian~\cite{luttinger}.
Of course, the standard Dresselhaus interaction~\cite{dresselhaus}
and even an additional Dresselhaus-like interaction~\cite{cardona}
exist for holes too.  They are relatively week in  semiconductors
with light atoms like GaAs, and they become more important in 
 semiconductors with heavy atoms like InAs and InSb~\cite{roland,Tommy}. 
Nevertheless, the most important spin-orbital
physics comes from the Luttinger Hamiltonian, and this is what we
consider in the present work.

Most, if not all, modern hole-based spintronics devises are essentially 
two-dimensional (2D). This includes quantum point contacts 
(QPC)~\cite{danneau,koduvayur,klochan,chen,komijani,ashwin1,nichele},
quantum dots~\cite{kulba,cress,klochdots,bulaev}, as well as heterostructures used in  Subnikov-de Haas 
measurements~\cite{Tommy,Yeoh,bangert,winkshj,failla}. These systems are based on quantum wells. The well
freezes dynamics in one direction due to quantum confinement. 
The confinement results in dimensional reduction of the Luttinger 
Hamiltonian, 3D$\to$2D, only the in-plane coordinate and 
associated in-plane momentum $\bm k$ remain as dynamic variables.
Due to interplay of the spin-orbit interaction,
the external magnetic field, and the lateral gate potential imposed on
the heterostructure spin dynamics of the arising 2D system is highly
nontrivial.
For example, for out-of-plane magnetic field the g-factor of hole
is significantly modified by the virtual 3D 
dynamics~\cite{wimbauer,durnev,drichko,simion}.
The virtual 3D dynamics is even more important for response
to the in-plane magnetic field~\cite{komijani,gradl}.
Effects of magnetic field in some 
dimensionally reduced systems have been considered 
previously, but only in some limiting 
cases~\cite{wimbauer,durnev,drichko,simion,andreani,komijani,gradl}.
The spin-orbital effects related to the lateral gate potential
to the best of our knowledge have been considered before
only in a very special limit related to the artificial
graphene~\cite{sushkov}.
In the present work we consider the most general situation with respect
to both the magnetic field and the lateral gate potential.
This analysis is applicable to QPCs, quantum dots, and for 
laterally modulated superlattices.

In the present work we derive the general two-dimensional effective
Hamiltonian resulting from the dimensional reduction of the
Luttinger Hamiltonian in a symmetric heterostructure.
We develop a general method for the dimensional
reduction valid for any symmetric heterostructure.
To be specific we present results of numerical calculations
for two different types of heterostructure,
(i) parabolic quantum well, (ii) infinite rectangular quantum well
in GaAs and InAs. 
In the present work we do not consider asymmetric heterostructures
which necessarily generate the Rashba-type effective interaction.
Such heterostructures require techniques beyond that
developed in the present work. Therefore, the asymmetric case will be a 
subject of a separate work.

The paper is organized as follows; 
in Section \ref{sbb} we briefly remind the very well-known calculation
of hole sub-bands, see e.g Ref.~\cite{roland}
We use results of this section in the rest of the paper.
In Section \ref{Heff} we introduce the
effective Hamiltonian as Ginzburg-Landau-type gradient expansion
over lateral potential. 
Section \ref{gxx} addresses momentum dependent in-plane g-factors.
In Section \ref{lat} we derive the spin-orbit interaction related
to the lateral potential.  To do so we use the scattering amplitude method
first developed for Breit interaction in quantum electrodynamics \cite{landau}. 
 Usage of the method allows us to find also the Darwin term in the effective
Hamiltonian.
Section \ref{gzzf} addresses momentum dependent out-of-plane g-factor
which is the most technically involved part. Finally, we present our conclusions
in Section \ref{conc}.

\section{Sub-bands} \label{sbb}
We start this section from reminding the very well known description
of 2D sub-bands which we use in subsequent sections.
Non-interacting holes in bulk conventional semiconductors are 
described by the Luttinger Hamiltonian~\cite{luttinger}.
In this paper, we consider so-called spherical approximation 
to the Hamiltonian
\begin{eqnarray}
H_L = \left(\gamma_1 + \frac{5}{2} \overline{\gamma}_2 \right)
\frac{{\bf p}^2}{2 m} 
- \frac{\overline{\gamma}_2}{m} \left({\bf p} \cdot {\bf S} \right)^2
\ ,
 \label{Lut1}
\end{eqnarray}
where~\cite{baldereshi}:
\begin{eqnarray}
\overline{\gamma}_2 = \frac{2 \gamma_2 + 3 \gamma_3}{5} \ .\nonumber
\end{eqnarray}
Here ${\bf p}$ is 3D quasi-momentum; ${\bf S}$ is the spin $S = 3/2$; 
$\gamma_1$, $\gamma_2$, and $\gamma_3$  are Luttinger parameters; 
and $m$ is the free electron mass.
It is known that
there is also a nonspherical  part of the  Luttinger Hamiltonian
\begin{eqnarray}
\delta H_L = p_ip_jS_mS_nT^{(4)}_{ijmn}\ ,
 \label{ns}
\end{eqnarray}
where the irreducible 4th rank tensor $T^{(4)}_{ijmn}$ depends on the 
orientation  of the cubic lattice. The tensor is proportional to 
$\gamma_2-\gamma_3$. Since in the large
spin-orbit splitting materials $\gamma_2 \approx \gamma_3$ the rotationally
noninvariant part of the Hamiltonian is small and we neglect it.
Of course, there are some effects that are essentially related to
the rotational asymmetry~\cite{winkler1,winkler2,Tommy,Yeoh} and
in this cases (\ref{ns}) cannot be neglected.
The general techniques developed in the present work can accommodate 
the rotational anisotropy. Nevertheless, for clarity of presentation
we omit the anisotropy in the present paper. 

\begin{figure}
\includegraphics[width=0.3\textwidth]{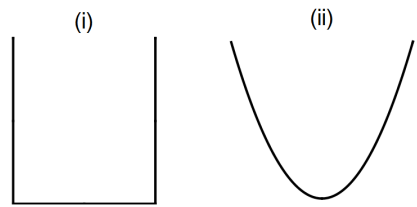}
\vspace{15pt}
\caption{Shape of quantum well:
(i) infinite rectangular, (ii) parabolic.
}
\label{shape}
\end{figure}

Impose the quantum well potential $W(z)$ on the system, the Hamiltonian
reads
\begin{eqnarray}
H=H_L +W(z) \ .
 \label{wz}
\end{eqnarray}
The well confines dynamics along z-axis leading to 2D sub-bands
$\varepsilon_{n,\bm k}$, where ${\bm k}=(k_x,k_y)=(p_x,p_y)$ is the 2D
momentum, and integer n enumerates the bands.
To be specific, we present herein numerical results
for parabolic and infinite rectangular quantum wells
in GaAs and InAs, see Fig.\ref{shape}.
\begin{eqnarray}
(i)&:& \ \ \ W(z)= \left\{
\begin{array}{cc}
0,& \quad z \in (-d/2, d/2) \\
\infty,& \quad {\rm otherwise}.
\end{array}
\right.
\nonumber\\
(ii)&:& \ \ \ W(z) = \frac{m \omega_z^2 z^2}{2}
\end{eqnarray}

\begin{figure}[h]
\includegraphics[width=0.225\textwidth]{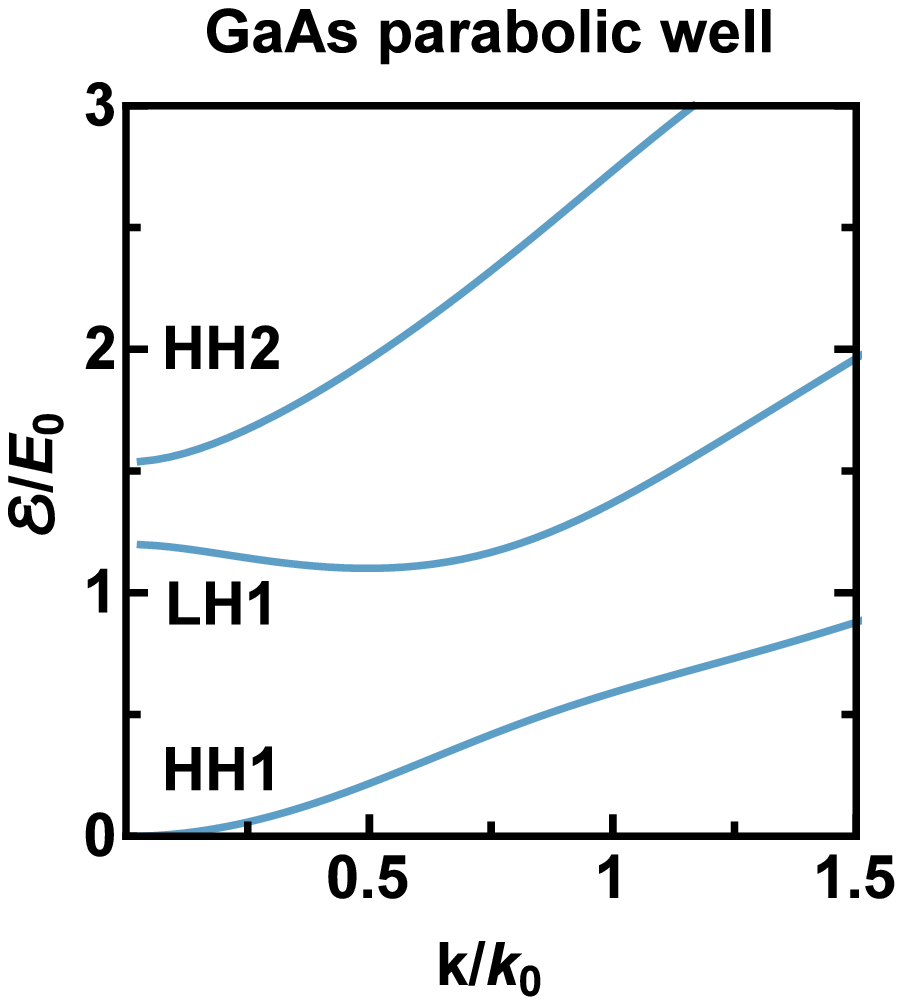}
\hspace{10pt}
\includegraphics[width=0.225\textwidth]{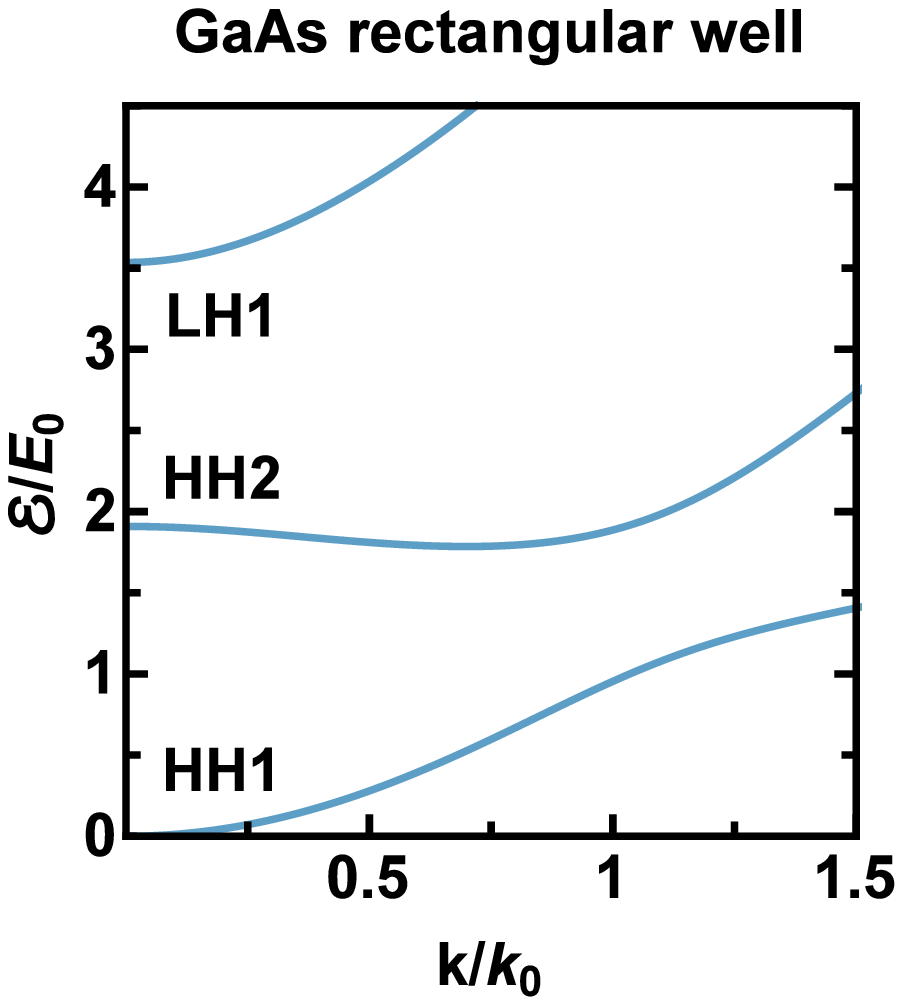}\\

\includegraphics[width=0.225\textwidth]{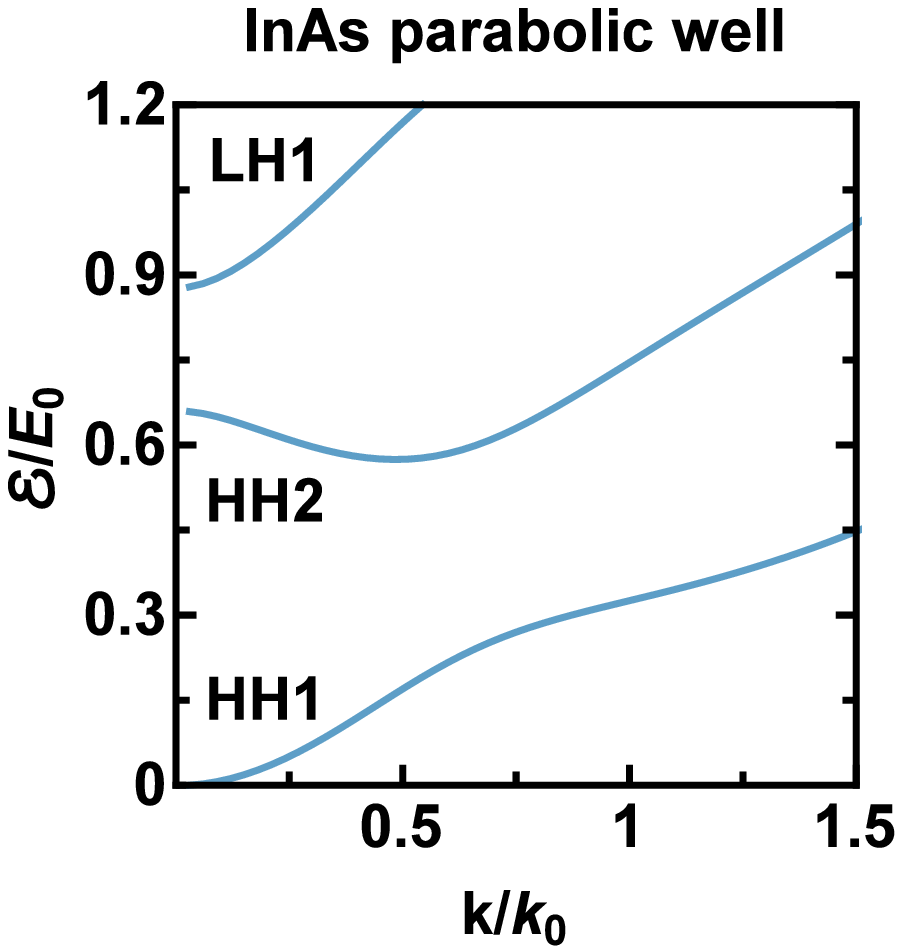}
\hspace{10pt}
\includegraphics[width=0.225\textwidth]{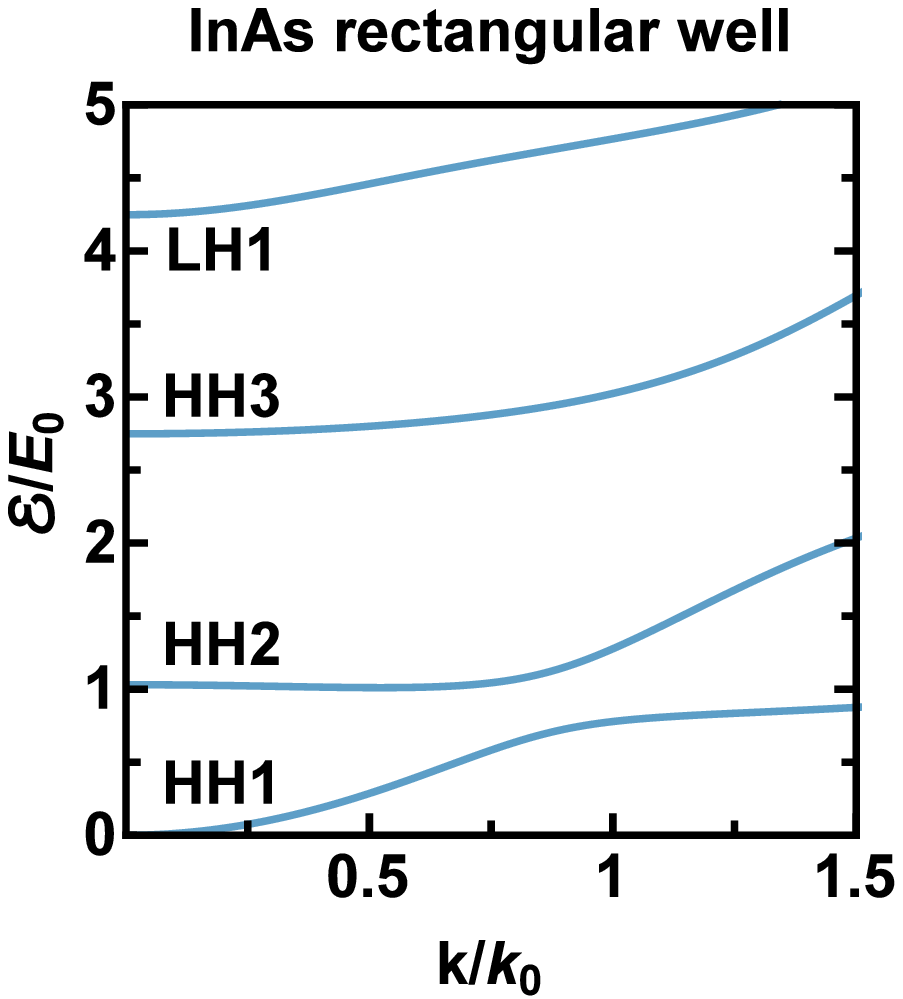}
	\caption{Hole sub-bands for  parabolic quantum well
and for infinite rectangular well in GaAs and InAs.
Momentum is given in units of $k_0$, see  Eq.(\ref{k00}), and
energy in units of $E_0$, see Eq.(\ref{e0}). 
}
	\label{spec}
\end{figure}

Since z-confinement is very strong, the most important part of (\ref{wz})
comes from terms proportional to $p_z^2$,
\begin{eqnarray}
H_0 &=&  \left( \gamma_1 + \frac{5}{2}\overline{\gamma}_2 -2 \overline{\gamma}_2
 S_z^2 \right) \frac{p_z^2}{2 m} +W(z)\nonumber\\
&+&\left( \gamma_1 - \frac{5}{4}\overline{\gamma}_2 +\overline{\gamma}_2
 S_z^2 \right) \frac{k^2}{2 m} \ .
\label{H0}
\end{eqnarray}
Note that the simple in-plane kinetic energy $\propto k^2/2m$
is also included in $H_0$.
It is evident from (\ref{H0}) that the lowest energy sub-band comes from
$S_z=\pm 3/2$.  It is usually called the first ``heavy hole''  subband (HH1).
There is also HH2 sub-band, etc.
The sub-bands with $S_z=\pm 1/2$ are called ``light hole'' sub-bands, 
LH1, LH2, etc. The Hamiltonian (\ref{wz}) can be represented as
\begin{eqnarray}
\label{h1}
&& H=H_0+V\\
&& V=-\frac{\overline{\gamma}_2}{4m}\left[k_+^2 S_-^2+k_-^2 S_+^2+ \right. \nonumber \\
&& \left. +\{p_z, k_+\} \{S_z, S_-\} +\{p_z, k_-\} \{S_z, S_+\}\right] \ .\nonumber
\end{eqnarray}
Here $\{...\}$ denotes the anticommutator.
Evaluation and diagonalization of the Hamiltonian matrix of (\ref{h1}) in the basis
of eigenstates of (\ref{H0}) is straightforward. The sub-bands arising
from this diagonalization for GaAs and InAs are plotted in Fig.\ref{spec} for both quantum wells  considered here.

\begin{figure}[h]
\includegraphics[width=0.49\linewidth]{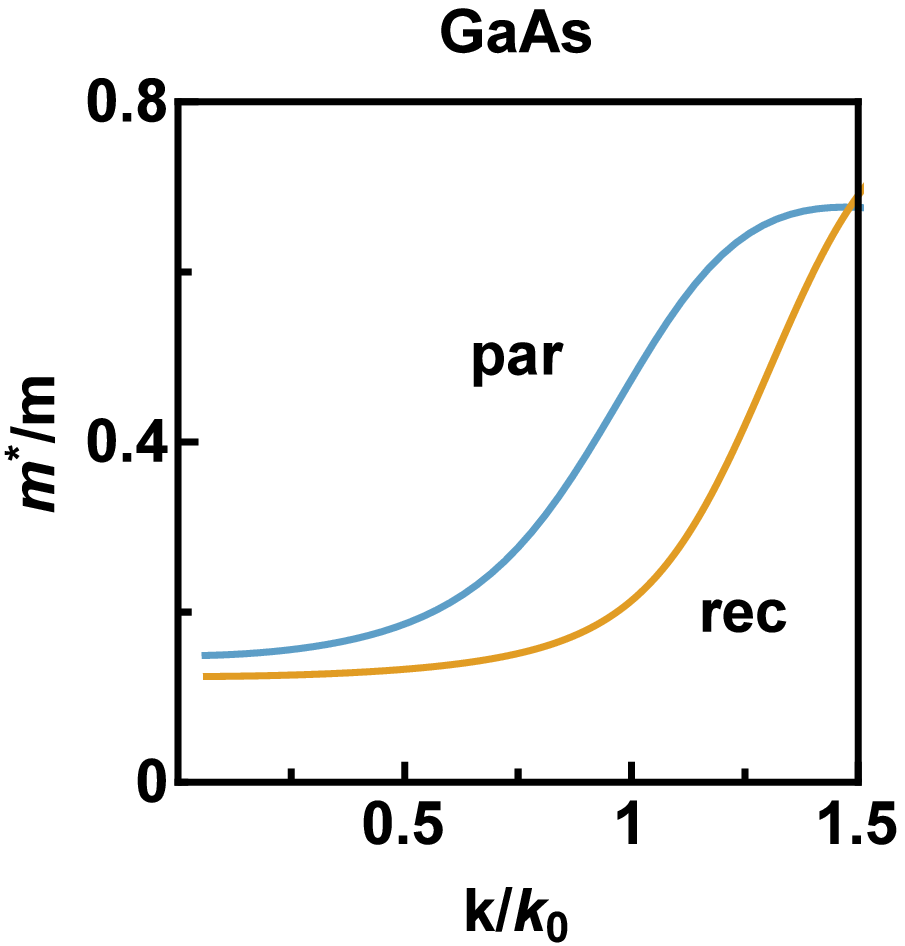}
\includegraphics[width=0.49\linewidth]{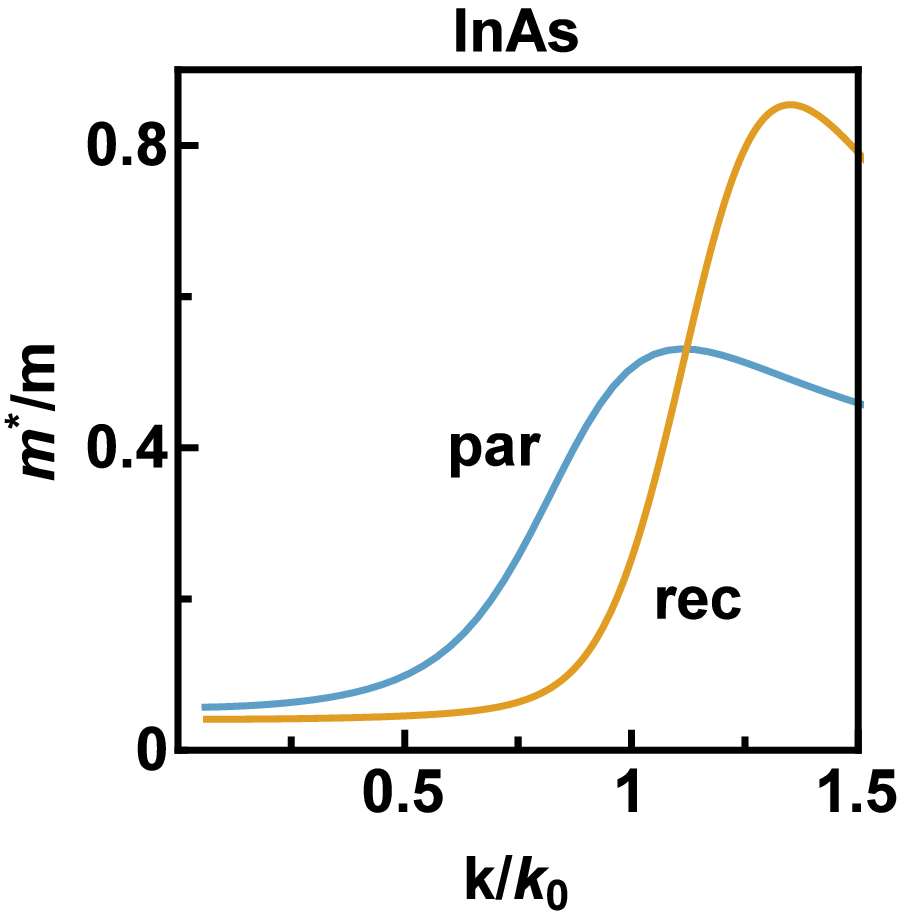}
\caption{The effective mass in the HH1 subband.
Parabolic and rectangular quantum wells in GaAs and InAs.
}
	\label{mass}
\end{figure}

Luttinger paramters used in the calculations 
are presented in Table I.
\begin{table}[h]
	\label{tab1}
	\begin{center}
		\begin{tabular}{|c|c|c|c|c|c|}
			\hline
  & $\gamma_1$ & $\gamma_2$ & $\gamma_3$ & ${\overline \gamma}_2$  &$\kappa$\\
			\hline
GaAs  &  6.85      &  2.1     &   2.9      & 2.58  & 1.2\\
\hline
InAs  & 20.4       & 8.3      &  9.1       & 8.78  &7.6\\
			\hline
			\end{tabular}
	\end{center}
	\caption{Luttinger parameters for GaAs, and InAs.}
\end{table}
Let us introduce the momentum unit for parabolic and rectangular wells:
\begin{eqnarray}
\label{k00}
k_0 = \left\{
\begin{array}{ll}
0.5 \sqrt{m \omega_z} & { \ \rm parabolic}\\
2.0/d                &  { \ \rm rectangular}
\end{array}
\right.
\end{eqnarray}
It is useful to note that for rectangular quantum well with $d=20$ nm,
the momentum $k=k_0$ corresponds to the hole density $1.6 \times 10^{11}cm^{-2}$,
this is about a typical experimental density.
Energies we express in units
\begin{equation}
\label{e0}
E_0 = \frac{\gamma_1 k_0^2}{2m}.
\end{equation}
The energy scale for rectangular quantum well of the width $d=20$ nm is $E_0 = 2.6$ meV which corresponds to experimental values of the Fermi energy. Effective mass $m^*=k\left(\frac{d\varepsilon}{dk}\right)^{-1}$ increases with the in-plane momentum due to non-parabolic corrections, see Fig.~\ref{mass}.

Let us impose now a magnetic field $B$ and a lateral potential $U(x,y)$
on the heterostructure. The magnetic field manifests itself in the
long derivatives in the Hamiltonian (\ref{H0}), (\ref{h1}), 
${\bm p}\to {\bm p}-e{\bm A}$, and in the Zeeman term 
$-2\kappa \mu_B ({\bm B}\cdot{\bm S})$. Here, ${\bm A}$ is the vector potential,
$e$ is the elementary charge, $\mu_B$ is the Bohr magneton. Values of the material specific parameter $\kappa$ are listed in the Table I, see e.g. Ref.~\cite{roland}
Thus, the total 3D Hamiltonian reads
\begin{eqnarray}
\label{h2}
&&H=H_0({\bm \pi})+V({\bm \pi}) -2\kappa \mu_B ({\bm B}\cdot{\bm S})
+U(x,y) \nonumber\\
&&{\bm \pi}={\bm p}-e{\bm A}
\end{eqnarray}

\section{Effective 2D Hamiltonian, gradient expansion} \label{Heff}
We consider here only HH1  sub-bands,
see Fig. \ref{spec}. This implies that the Fermi energy is below the
bottom of the first excited sub-band.
The HH1  subband is double degenerate due to the Kramers theorem.
The standard way to describe
the Kramers doublet is to introduce the effective spin $s=1/2$ (pseudo-spin) with
related Pauli matrixes ${\bm \sigma}$. The correspondence at $k=0$
is very simple, $|\uparrow\rangle=|S_z=3/2\rangle$,
$|\downarrow\rangle=|S_z=-3/2\rangle$. 

The effective 2D Hamiltonian can depend only on 2D variables,
it cannot contain $z$ and $p_z$. We assume that the gate lateral
potential is smooth, $k\nabla U \ll U$, so we can use the gradient expansion 
of the potential. As soon as we understand this point, the kinematic form of the
Hamiltonian is unambiguously dictated by symmetries and gauge invariance:
\begin{eqnarray}
\label{heff}
H_{2D}&=&\varepsilon({\bm \pi}) +U(x, y)\nonumber\\
&+&\{\alpha({\bm \pi}),({\bm s}\cdot 
[{\bm \nabla} U \times{\bm \pi}])\}
+\frac{1}{2}\{\beta({\bm \pi}),\Delta U\} \nonumber\\
&-&g_{zz}({\bm \pi}) \mu_B B_z s_z\nonumber\\
&-&\frac{\mu_B}{2}{\overline g_{1}}({\bm \pi})(B_+\pi_+^2 s_- + B_-\pi_-^2 s_+)\nonumber\\
&-&\frac{\mu_B}{2}{\overline  g_{2}}({\bm \pi})(B_-\pi_+^4s_- + B_+\pi_-^4 s_+)  .
\end{eqnarray}
Here, $\textbf{s} = {\bm \sigma}/2$ is pseudo-spin; $B_{\pm}$, $\pi_{\pm}$, and  $s_{\pm}$ are defined in the
standard way, $B_{\pm}=B_x\pm iB_y$ etc, $\{ \dots \}$ is anti-commutator.
We stress again that all the variables, gradients, etc.
in the Hamiltonian are two dimensional.
While the dispersion $\varepsilon({k})$ has been discussed in the previous 
section, the functions $\alpha({k})$, $\beta({k})$, $g_{zz}({k})$,
${\overline  g_{1}}({k})$, and ${\overline  g_{2}}({k})$ 
will be determined in subsequent
sections. Like the dispersion they are isotropic, i.e. depend on $k=|{\bm k}|$.
It is useful to underline the most important points concerning Eq.~(\ref{heff}).
(i) This is a gradient expansion up to the second gradient of $U$,
we neglect third derivatives of the gate potential.
The Darwin term, $\propto \Delta U$, is spin independent and therefore
not very interesting. We keep it only for completeness, since like in
Dirac equation it comes together with the usual spin-orbit
and the coefficients $\alpha$ and $\beta$ have the same dimension.
(ii) Due to the gauge invariance, all the coefficients, $\alpha$, $\beta$, 
$g_{zz}$, ${\overline g_{1}}$, and ${\overline g_{2}}$
depend on the kinematic momentum ${\bm \pi}={\bm k}-e{\bm A}$, ${\bm A}$ depends only on x and y. 
(iii) The $\alpha$- and $\beta$-terms contain anticommutators.
(iv) We neglect powers of magnetic field higher than two in the spin
response, for example, we neglect the kinematic structures like
$B_+^3\sigma_-$. However, it is very easy to restore these terms with the developed technique.
(v) The angular momentum selection rule for the effective spin $\sigma_-$
is $\Delta S_z =-3$. Therefore, there are two generally independent
in-plane g-factors, the  functions ${\overline  g_1}$ and ${\overline  g_2}$.
(vi) The Hamiltonian must be symmetric over all rotations around z-axis and $\pi$-rotations around x-axis.
 Therefore, ${\overline g_{1}}$ and ${\overline g_2}$ are real. Of course, $\alpha$, $\beta$  and $g_{zz}$ 
are real too.

Now we proceed to the calculation of functions which enter the effective 
Hamiltonian (\ref{heff}), we start from the in-plane g-factors.

\section{in-plane g-factors} \label{gxx}
\begin{figure}[h]
\includegraphics[width=0.22\textwidth]{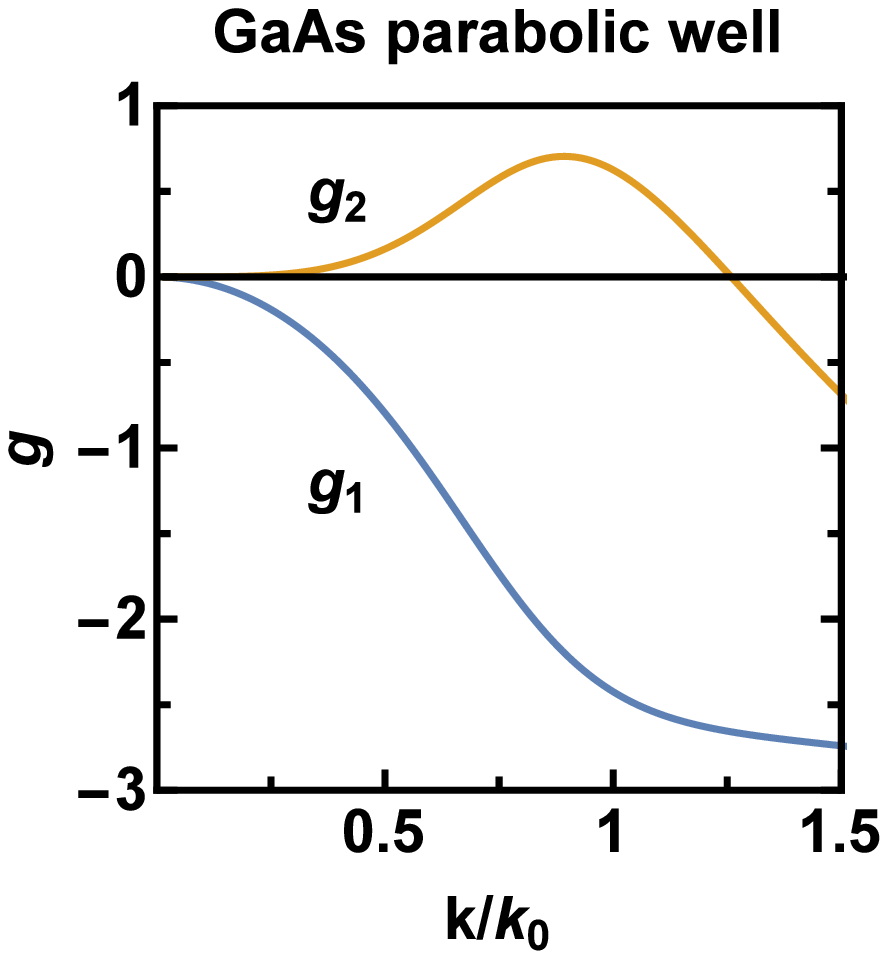}
\hspace{10pt}
\includegraphics[width=0.22\textwidth]{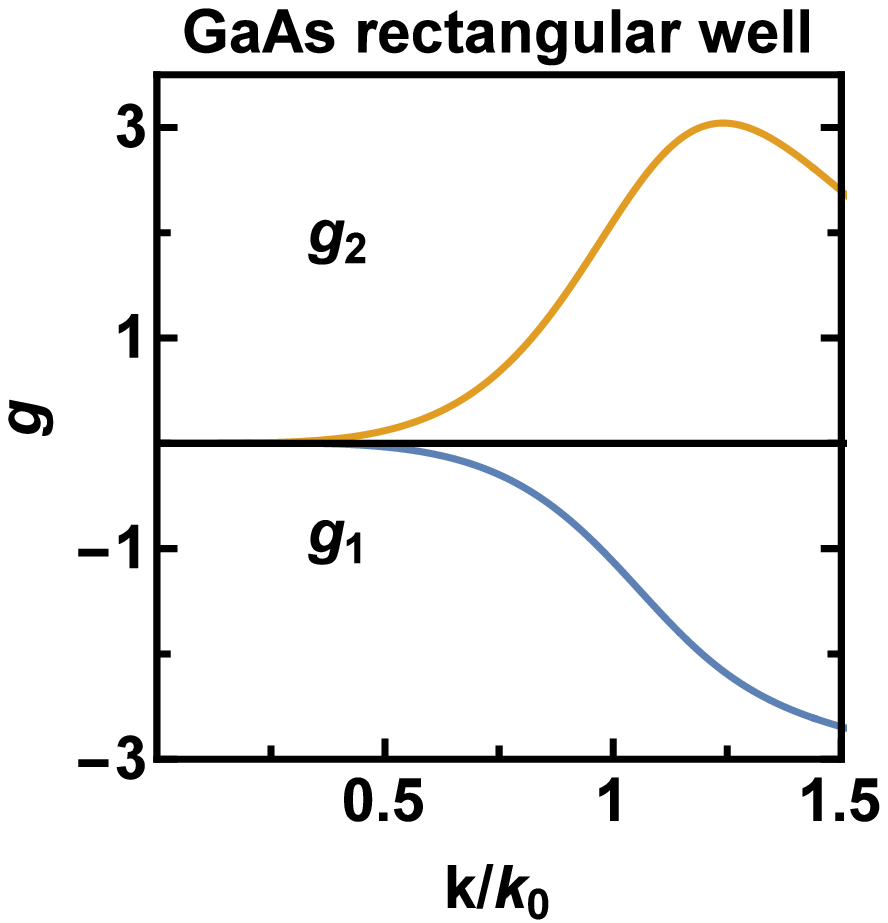}

\includegraphics[width=0.22\textwidth]{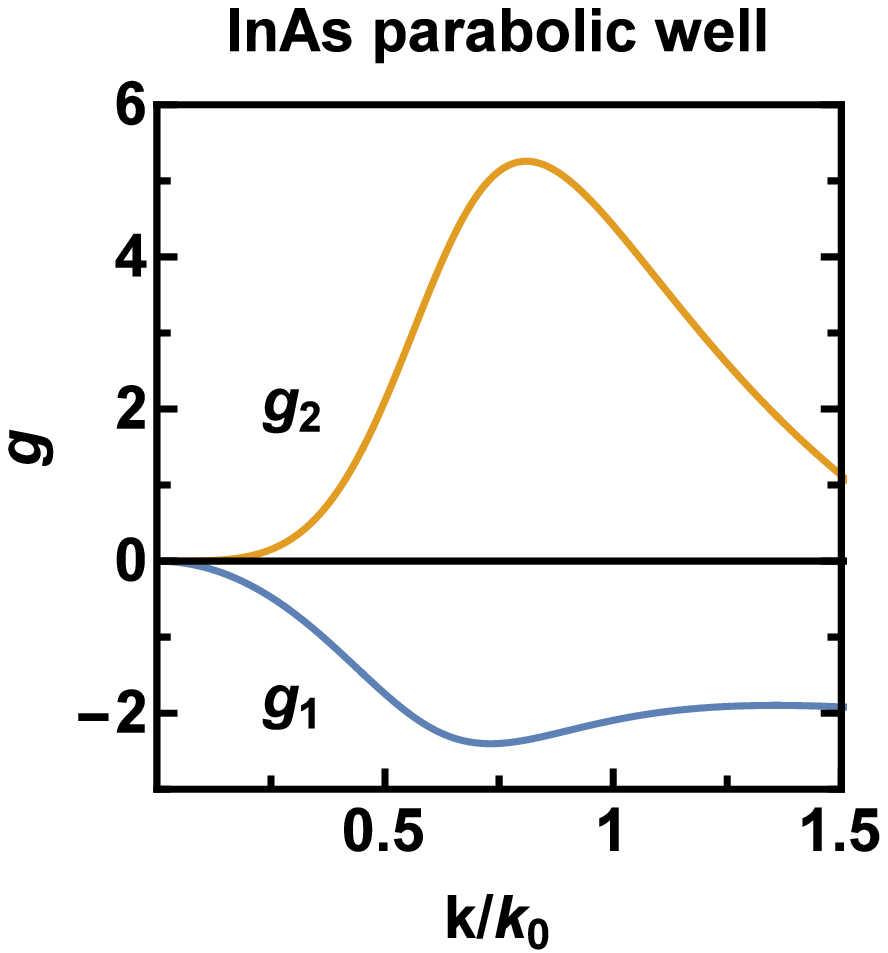}
\hspace{10pt}
\includegraphics[width=0.22\textwidth]{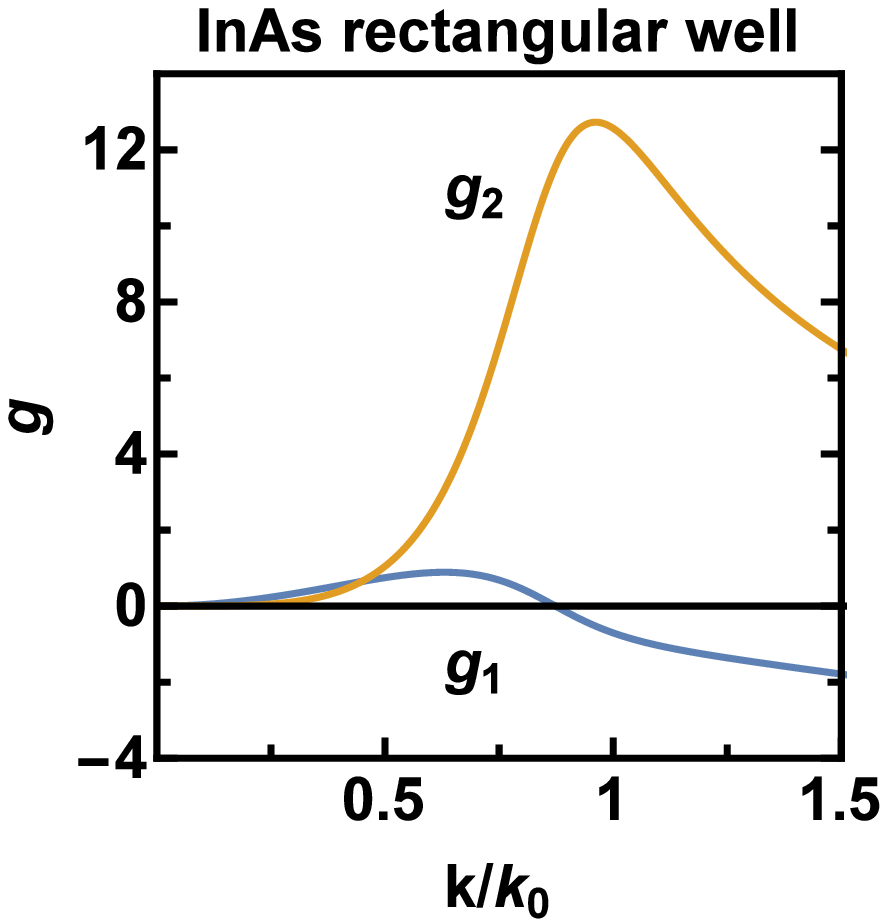}
	\caption{Effective in-plane HH1 g-factors, see Eq.(\ref{g12}),
 versus momentum. The g-factors are presented for parabolic and
rectangular confinement in GaAs and InAs.
}
\label{nu}
\end{figure}

There are two independent in-plane g-functions, ${\overline  g_1}(k)$ 
and ${\overline g_2}(k)$, see Eq. (\ref{heff}).
In order to find these functions, we numerically diagonalize the Luttinger 
Hamiltonian (\ref{h2}) with $U=0$ and with
magnetic field directed along x-axis, ${\bm B}=(B,0,0)$.
We use the vector potential in the following gauge:
\begin{equation}
{\bf A} = (0, -B z, 0).
\end{equation}
In this gauge,  the in-plane momentum ${\bm k}$ is a good quantum number,
$\psi \propto e^{ik_xx+ik_yy}$. Therefore ${\bm k}$ enters  (\ref{h2})
as a simple number, so only the one-dimensional z-confinement problem
needs to be diagonalized numerically.
We calculate the magnetic spin-splitting of HH1 sub-band 
at different values of ${\bm k}$. 
As we have to find two different functions, we perform calculation twice 
with ${\bm k}=(k,0)$ and with ${\bm k}=(0,k)$. 
Effective momentum dependent g-factors
\begin{eqnarray}
\label{g12}
&&g_{1} = k^2  {\overline g_1}(k)\nonumber\\
&&g_{2} = k^4  {\overline g_2}(k)
\end{eqnarray}
for GaAs and InAs and for parabololic and rectangular quantum wells
are plotted in Fig.\ref{nu}. 

\begin{figure}[h]
\includegraphics[width=0.22\textwidth]{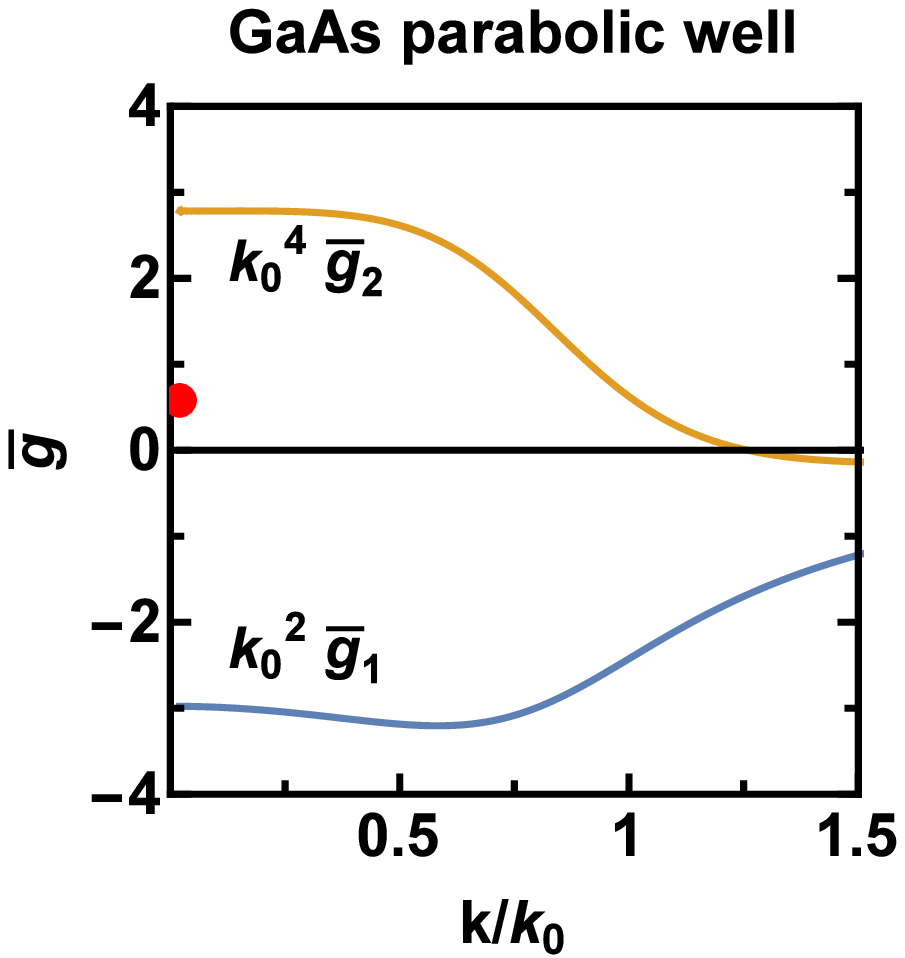}
\hspace{10pt}
\includegraphics[width=0.22\textwidth]{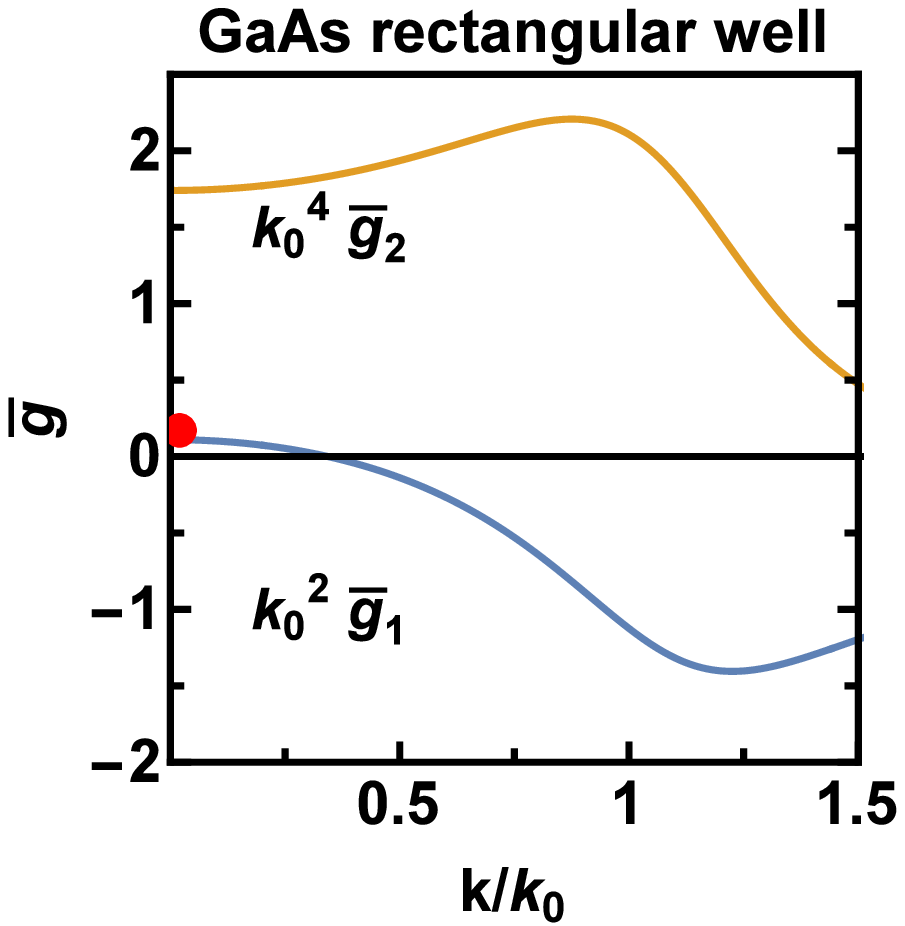}

\includegraphics[width=0.23\textwidth]{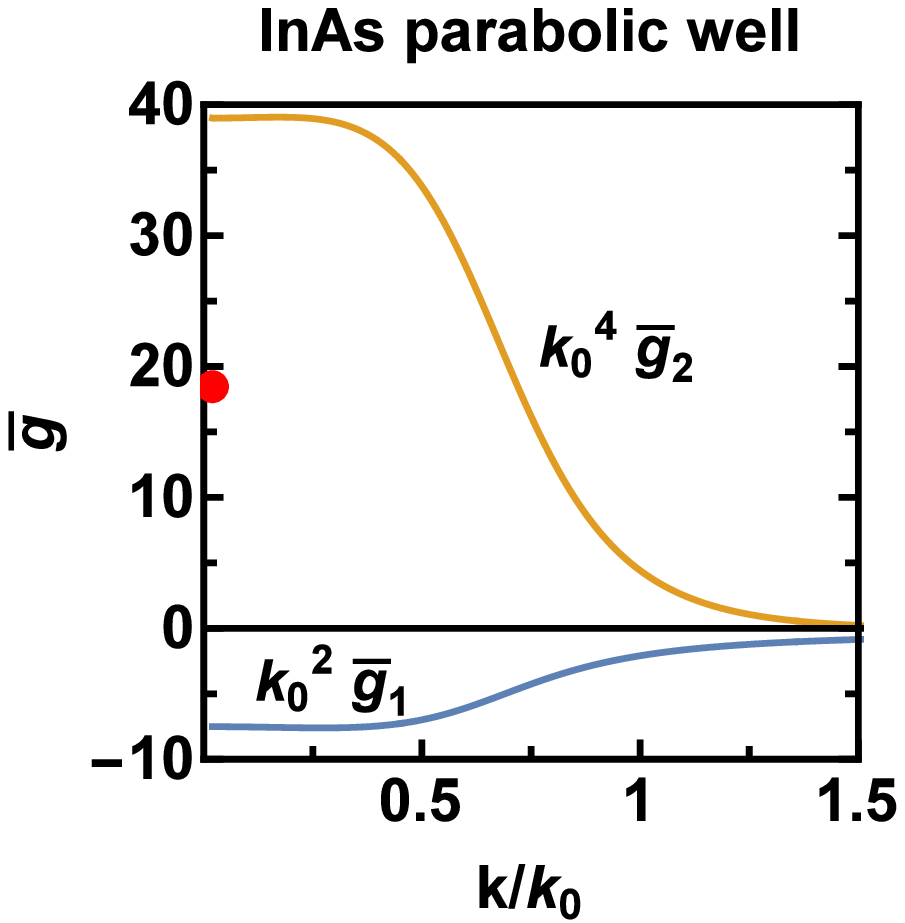}
\hspace{10pt}
\includegraphics[width=0.22\textwidth]{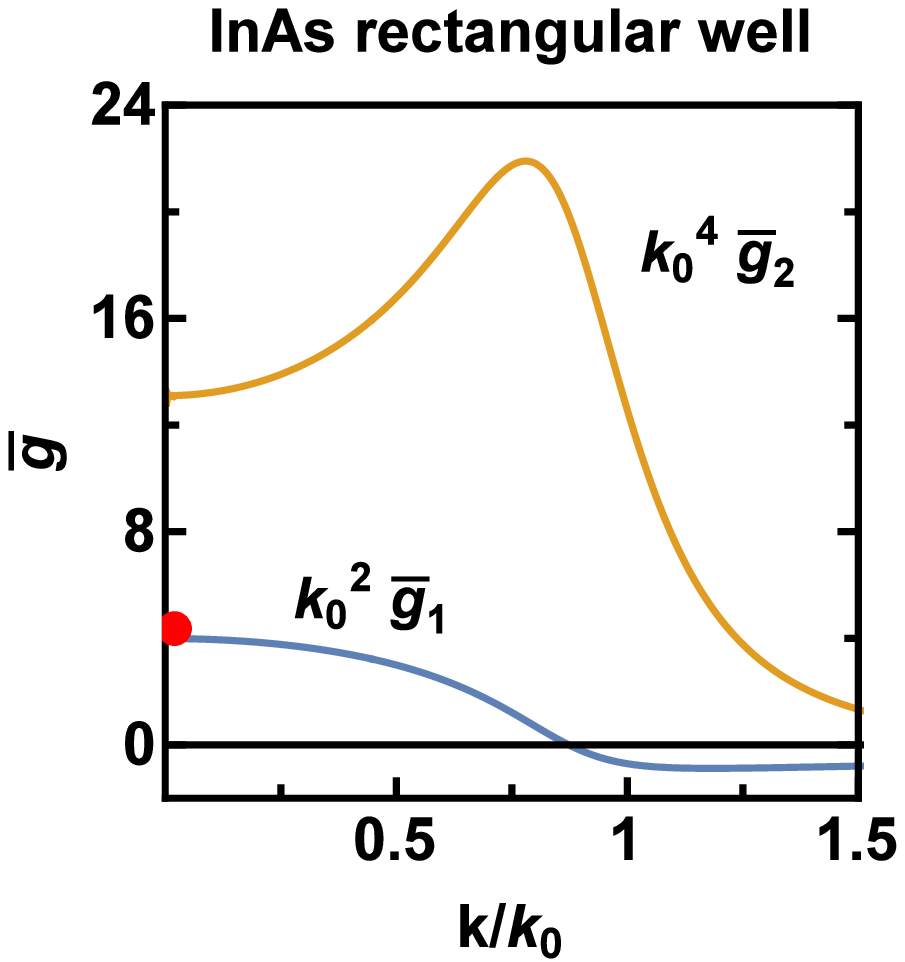}
	\caption{Functions ${\overline g_1}(k)$ and ${\overline g_2}(k)$
related to the in-plane g-factors, see Eq.~(\ref{g12}). 
The functions are presented for parabolic and rectangular quantum wells 
in GaAs and InAs.
To balance dimension,
we plot the functions multiplied by a corresponding power of $k_0$.
Red points on the vertical axes indicate values of ${\overline g_1}(k=0)$
calculated using equations in Ref.~\cite{komijani} The points must be compared with
our blue lines.
}
\label{beta}
\end{figure}

At small $k$, the g-factors (\ref{g12}) scale as high powers of momentum.
Therefore, it is instructive to plot also  ${\overline g_1}$
and ${\overline g_2}$. These functions have dimensions
$[1/k^2]$ and $[1/k^4]$ respectively.
 We use powers of $k_0$, Eq.(\ref{k00}), to balance the momentum 
dimension. Plots of $k_0^2{\overline g_1}(k)$ and 
$k_0^4{\overline g_2}(k)$  are presented in Fig.\ref{beta}.
Zero momentum value of ${\overline g_1}$ can be calculated analytically
with usual perturbation theory, the result reads
\begin{eqnarray}
\label{MZN}
&&{\overline g_1}(0) = -3 {\overline{\gamma}}\left(\kappa Z_1 - 4 {\overline{\gamma}} Z_2 -{\overline{\gamma}} Z_3 + 2 \kappa {\overline{\gamma}} Z_4 \right) \\
&&Z_1 = - 2 \sum\limits_{n=1}^\infty \frac{|\langle 1H | nL \rangle|^2}
{m(\varepsilon_{nL} - \varepsilon_{1H})} \nonumber \\
&&Z_2 = 2 i \sum\limits_{n = 1}^\infty \frac{\langle 1H| z |nL \rangle 
\langle nL |p_z| 1H \rangle}{m(\varepsilon_{nL} - \varepsilon_{1H})} \nonumber\\
&& Z_3 = -2 i \sum\limits_{n = 1}^\infty \frac{\langle 1H | \{z, p_z\} |n L \rangle \langle n L | 1H \rangle}{m(\varepsilon_{nL} - \varepsilon_{1H})} \nonumber \\
&&Z_4 = 2 \sum\limits_{n=1}^\infty \frac{|\langle 1H | p_z | nL \rangle|^2}{m^2(\varepsilon_{nL} - \varepsilon_{1H})^2}\ .\nonumber
\end{eqnarray}
The zero momentum value of ${\overline g_1}$ has been calculated in 
Ref.~\cite{komijani} with only $Z_1$ and $Z_2$ terms taken into
account. 
$Z_4$ and, especially, $Z_3$-term are important, this is why our values
of ${\overline g_1}(0)$ differ from that of Ref.~\cite{komijani} 
shown in Fig.\ref{beta} by red dots on the vertical axis.

The most important conclusion of this section is that at $k \approx 1$
where most experiments are performed, both invariant g-factors
$g_1$ and $g_2$ are equally important, see Fig.~\ref{nu}. 

\section{spin-orbit interaction due to the lateral gate potential}
\label{lat}
To calculate the spin-orbit interaction and the Darwin term
in the effective Hamiltonian (\ref{heff}) we employ the
scattering amplitude method which is usually used for
derivation of Breit interaction in quantum electrodynamics \cite{landau}.
This is a technically efficient way to proceed from a full multicomponent
description to the effective two-component wave function.
Magnetic field is not relevant to this problem, so in this section
the magnetic field is zero.
Consider scattering of a hole from a weak lateral potential limited in space, 
for example, from a Gaussian potential,
\begin{equation}
\label{gau}  
U(x,y)=U_0e^{-(x^2+y^2)/a^2} \ .
\end{equation}
Actual shape of the potential is not important, the only important point
is that the potential is weak and limited in space, so the scattering problem 
makes sense.
The idea of the method is to calculate the Born scattering amplitude,
${\bm k} \to {\bm k}'$. The scattering amplitude calculated
with the effective Hamiltonian (\ref{heff}) must be the same as the
amplitude calculated with 3D Hamiltonian (\ref{h2}). This allows one to
find functions $\alpha(k)$ and $\beta(k)$ in (\ref{heff}).  

An eigenstate of the Hamiltonian (\ref{H0}), $|S_z,n,{\bm k}\rangle$,
 possess a definite value of the in-plane
momentum ${\bm k}$ and a definite value of $S_z=-3/2,-1/2,1/2,3/2$,
\begin{equation}
\label{psi0}
|S_z,n,{\bm k}\rangle= e^{i{\bm k}\cdot{\bm r}}
|S_z,n\rangle \ .
\end{equation}
Here, the index n enumerates transverse modes (z-standing waves).
The diagonalization of the Hamiltonian (\ref{h1}) which is described in the section
\ref{sbb} results in energy bands and in wave functions
expressed in terms of (\ref{psi0}).
In particular, the wave function of the $|\uparrow,{\bm k}\rangle$ state of the
HH1 band is of the form
\begin{equation}
\label{psi1}
|\uparrow,{\bm k}\rangle=
 \sum_{S_z} \sum_n a_n(S_z,k) k_+^{(3/2-S_z)}|S_z,n,{\bm k}\rangle \ ,
\end{equation}
where the momentum dependent coefficients $a_n(S_z,k)$ are determined
by the numerical diagonalization, and the phase factor, $k_+^{(3/2-S_z)}$
is dictated by the conservation of total angular momentum.
The Born scattering amplitude is given by the matrix element of the 
scattering potential $U$ between the initial and final states,
\begin{eqnarray}
\label{kernel}
f_{{\bf k'} {\bf k}} = \langle \uparrow,{\bm k}'| U({\bf r})| \uparrow,{\bm k}
 \rangle=U_{\bf q} \sum_{l = 0}^3  b_l(k) \cdot (k'_- k_+)^l \ ,
\end{eqnarray}
where
\begin{eqnarray}
\label{bb}
b_l(k) = \sum\limits_{n = 1}^\infty \left|a_n(\frac{3}{2}-l,k)\right|^2\ ,
\end{eqnarray}
 $U_{\bm q}$ is the Fourier transform of  $U({\bf r})$, and 
${\bm q}={\bf k'}-{\bf k}$ is the momentum transfer.
When calculating (\ref{kernel}) using (\ref{psi1}), we take into account
that ${\bf k'}^2={\bf k}^2$ due to the energy conservation.
Note that the wave function normalization condition reads
\begin{equation}
 1 = b_0+ k^2 b_1  + k^4 b_2 + k^6 b_3\ .
\label{norm0}
\end{equation}
The product $k'_- k_+$ which enter Eq.~(\ref{kernel}) reads
\begin{equation}
\label{pro}
k'_- k_+=k^2-\frac{1}{2}q^2+i[q_xk_y-q_yk_x] \ .
\end{equation}
The first power of the square bracket in this equation is responsible
for the skew scattering. Therefore, comparing the $[{\bm q}\times {\bm k}]$
term from (\ref{kernel}) with
the scattering amplitude calculated with the $\alpha$-term in Eq.~(\ref{heff})
we find the following expression for $\alpha$
\begin{equation}
\label{ak}
\alpha(k) = b_1(k) + 2 k^2  b_2(k) + 3 k^4 b_3(k)  \ .
\end{equation}
Similarly, the terms proportional to $q^2$ in (\ref{kernel})
contribute to the Darwin term. Comparing  the $q^2$ term from (\ref{kernel}) 
with the scattering amplitude calculated with the $\beta$-term in 
Eq.~(\ref{heff}) we find the following expression for $\beta$
\begin{equation}
\label{bk}
\beta(k) = \frac{1}{2}b_1(k) +2 k^2 b_2(k) + \frac{9}{2} k^4 b_3(k) \ .
\end{equation}
There are also higher powers of $q$ in Eq.~(\ref{kernel}), up to $q^6$,
they correspond to higher gradients in the gradient expansion of the
effective Hamiltonian. We neglect the higher gradients in (\ref{heff})
assuming that $U$ is sufficiently smooth. However, in principle,
the scattering amplitude method Eq.~(\ref{kernel}) allows to find all the terms of the 
gradient expansion.

\begin{figure}[h]
\includegraphics[width=0.22\textwidth]{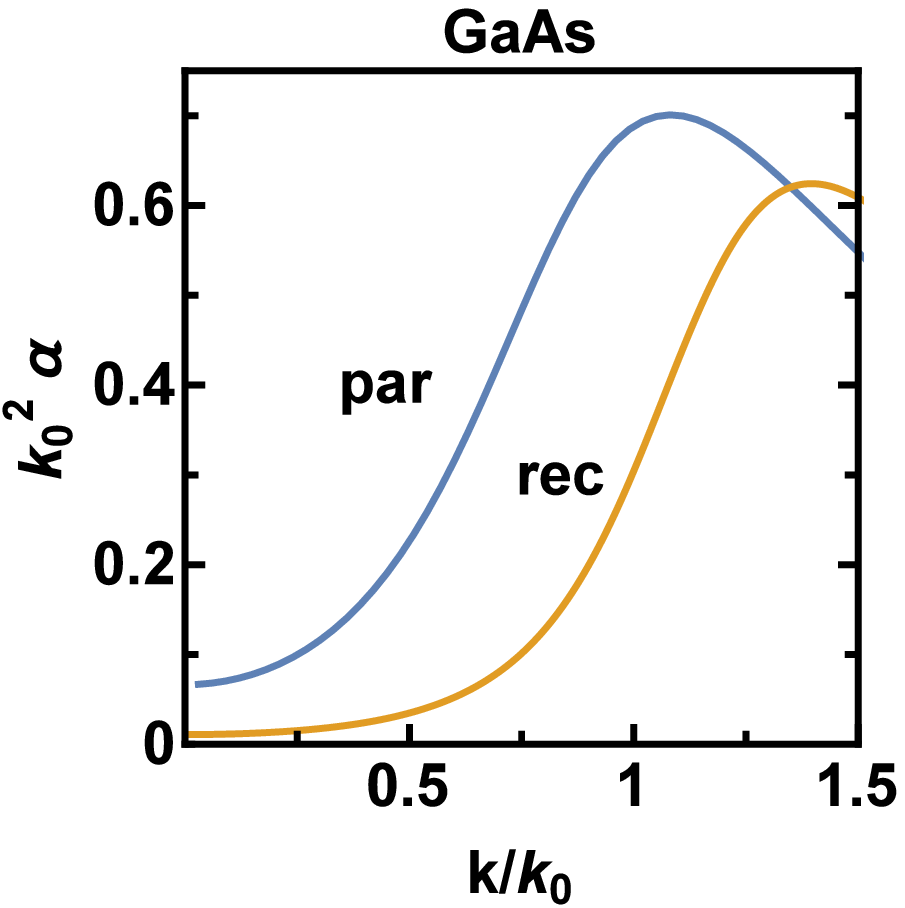}
\includegraphics[width=0.22\textwidth]{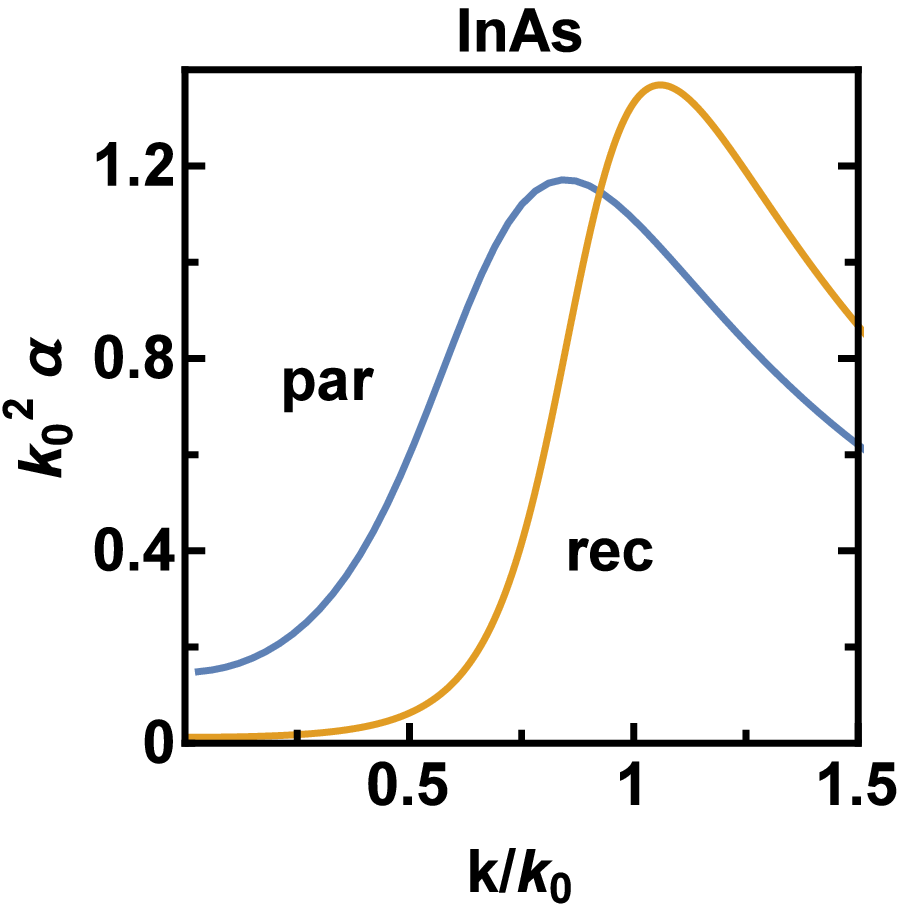}
\includegraphics[width=0.22\textwidth]{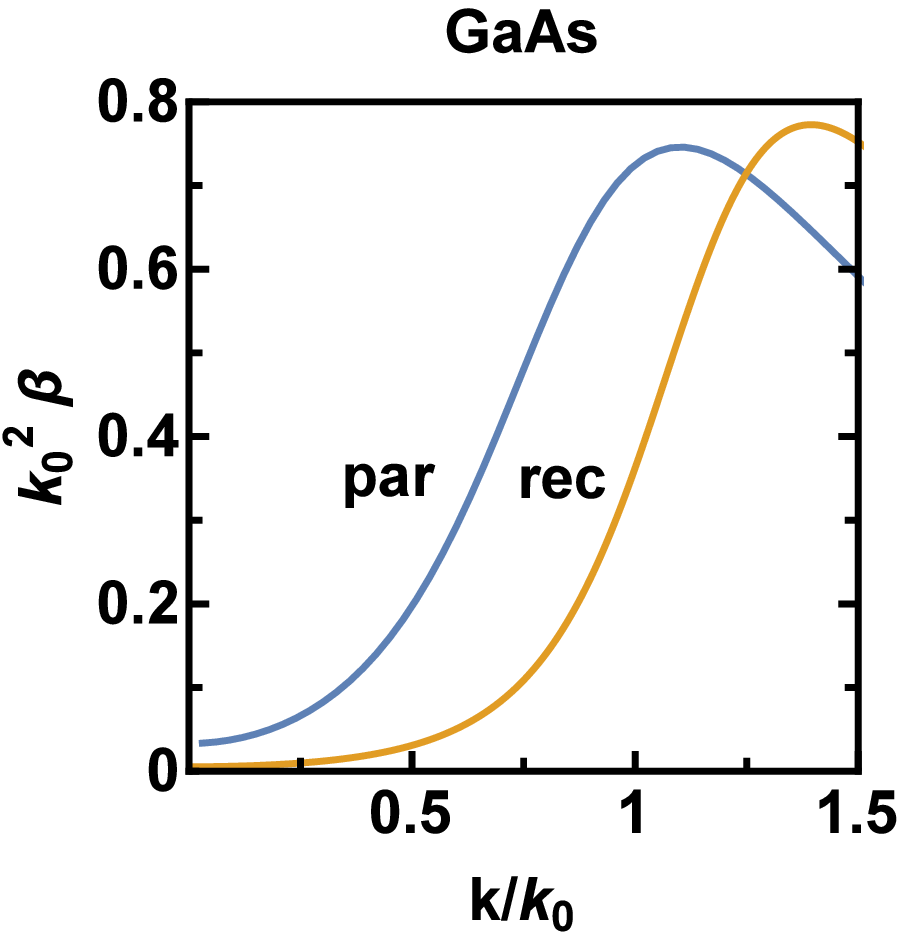}
\includegraphics[width=0.22\textwidth]{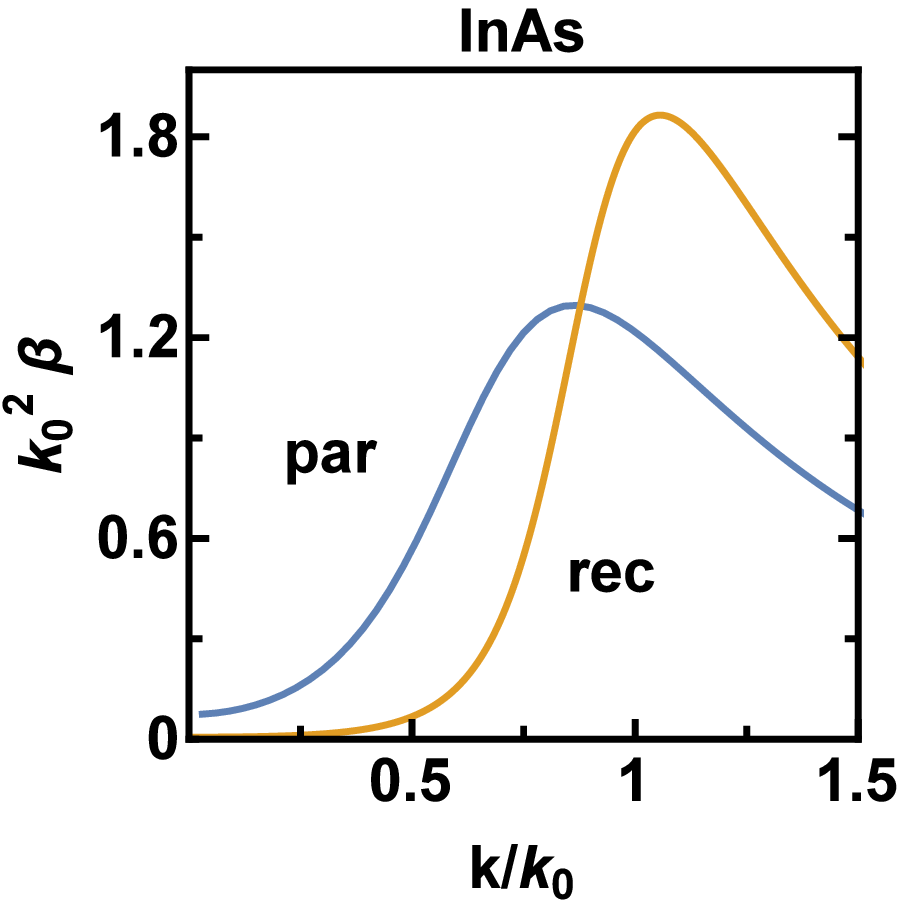}
	\caption{The spin-orbit $\alpha$ and the Darwin $\beta$ functions 
for the lateral gate potential, see Eq.~(\ref{heff}).
The functions are presented for parabolic and rectangular quantum wells 
in GaAs and InAs and multiplied by powers of $k_0$ to make them
dimensionless.
}
	\label{SOI}
\end{figure}

Since the numerical diagonalization described in section \ref{sbb}
gives all coefficients of the HH1 wave function, it is easy to calculate
$\alpha$ and $\beta$ using Eqs.~(\ref{bb}),(\ref{ak}),(\ref{bk}).
Functions $\alpha(k)$ and $\beta(k)$ for parabolic and  rectangular 
quantum wells in GaAs and InAs are plotted in Fig.~\ref{SOI}.
Both $\alpha$ and $\beta$ have dimension $[1/k^2]$.
To balance the dimension in Fig.~\ref{SOI}, we plot $k_0^2\alpha$
and $k_0^2\beta$.

\section{Out-of-plane g-factor}\label{gzzf}
A naive value of $g_{zz}$ is $g_{zz}=6 \kappa$, see Eq.~(\ref{h2}) and
Ref.~\cite{roland}
Virtual orbital 3D dynamics strongly influences this value. The effect of the
virtual dynamics at k=0
has been calculated previously, it leads to a very significant reduction
of the g-factor~\cite{wimbauer,durnev,drichko,simion}
\begin{eqnarray}
g_{zz}(0) = 6 \kappa - 12 {\overline \gamma}^2 \sum_{n = 1}^\infty \frac{|\langle 1 H | p_z | n L \rangle|^2}
{m(\varepsilon_{nL} - \varepsilon_{1H})}\ ,
\label{S} 
\end{eqnarray}
The g-factor is significantly different from the naive value.
For example, the g-factor in GaAs where $6\kappa=7.2$, is $g_{zz}(0)=7.2-5.15=2.05$ for parabolic quantum well
and $g_{zz}(0)=7.2-2.6=4.6$ for rectangular quantum well.
Our goal in this section is to calculate the entire function $g_{zz}(k)$
defined in Eq.~(\ref{heff}).
One possibility is to calculate Landau levels with the Hamiltonian
(\ref{h2}) and then look at the spin splitting of the levels.
This approach used in Ref.~\cite{simion} for the rectangular well indicated a significant dependence of g-factor on Landau level.
However, this method is rather technically involved and computationally
expensive. Here we use a different method, we destroy Landau levels
by a gate potential and calculate linear spin response.

\begin{figure}[h]
\includegraphics[width=0.2\textwidth]{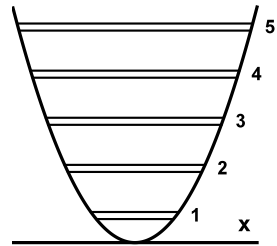}
	\caption{Magnetic field split oscillator levels with
$n_x=1,2,3,...$ in the parabolic gate potential (\ref{gp0}).
}
	\label{lev}
\end{figure}

Let us consider the parabolic gate potential,
\begin{equation}
\label{gp0}
U(x) = \frac{m \omega_x^2 x^2}{2} \ ,
\end{equation}
which restricts the hole propagation in the x-direction.
The vector potential for the out-of-plane magnetic filed ${\bm B}=(0,0,B)$  
is taken in the following gauge:
\begin{equation}
\label{aa}
{\bf A} = (0, B \cdot x, 0) \ .
\end{equation}
So, the y-component of momentum is conserved and we set $k_y=0$.
In this situation, the long momentum that enters the full Luttinger
Hamiltonian (\ref{h2}) is ${\bm \pi}=(p_x,-eBx,p_z)$.
Since there is no dynamics along the y-direction, $e^{ik_yy}=1$,
effectively the Hamiltonian (\ref{h2}) becomes two-dimensional,
only the x- and the z-directions are nontrivial.
A brute force numerical diagonalization of this Hamiltonian is straightforward.
We consider energy below the bottom of the first excited band, 
see Fig.~\ref{spec}, so all quantum states we consider
originate from the lowest HH1 band.
Due to the gate confinement (\ref{gp0}), the  spectrum is discrete, it is
described by an integer quantum number $n_x=1,2,3...$, and due to the
magnetic field, the Kramers degeneracy of each $n_x$ level is lifted as it 
is illustrated in Fig.~\ref{lev}.
For numerical calculations with (\ref{h2}),(\ref{gp0})
we use sufficiently small values of $\omega_x$, $\omega_x\sim 0.01$, so that 
there are
about 100-150 oscillator levels within the energy span of the HH1 band.
It is worth noting that the spectrum is not equidistant
because the dispersion $\varepsilon(p)$ is not parabolic.
Magnetic field which we consider in this analysis is even smaller than 
$\omega_x$,  $B \ll \omega_x$. In practice, we first diagonalize
(\ref{h2}),(\ref{gp0}) numerically at $B=0$, and then we account
for the weak magnetic field by usual perturbation theory.
As result, we find the Zeeman splitting of each oscillator level, see
Fig.~\ref{lev},
\begin{equation}
\label{gex}
\Delta \varepsilon=- g(\varepsilon)\frac{1}{2} \mu_B B \sigma_z\ .
\end{equation}
Here, $\varepsilon$ is energy corresponding to the oscillator level
with quantum number $n_x$. 
The outcome of the brute force numerical calculation
is the function $g(\varepsilon)$.

The next question is: how to deduce the function $g_{zz}(k)$ in Eq.(\ref{heff})
from 
$g(\varepsilon)$ obtained in the numerical calculation described
in the previous paragraph?
To answer this question,  let us articulate the problem in terms
of the effective 2D Hamiltonian (\ref{heff}).
With gate potential (\ref{gp0}) the effective Hamiltonian takes the 
following form
\begin{eqnarray}
H_{2D} &=& \varepsilon(k_x) + \frac{m \omega_x^2}{2} (x^2 + 2\beta) \nonumber \\
&-&[g_{zz} + 2 m^2 \omega_x^2 \{\alpha, x^2\}]\frac{1}{2} \mu_B B \sigma_z \ .
\label{eff}
\end{eqnarray}
Actually, the Hamiltonian becomes 1D.
We remind that $\alpha(k_x)$ is the coefficient in the
gate spin orbit interaction
and $\beta(k_x)$ is the coefficient in the
gate Darwin term, both functions have been 
determined in section \ref{lat}. The $\alpha$-term in (\ref{eff}) is due 
to $\pi_y=-eBx$.
Eq. (\ref{eff}) is written in linear in B approximation,
we neglect $\pi_y^2=(eBx)^2$ which is quadratic in B.
 There is an important point to note about the Hamiltonian (\ref{eff}):
the Zeeman splitting arises not only due to $g_{zz}$, there is a part
of the splitting which is due to the gate potential.
This point is important for understanding of experiments with quantum point 
contacts and quantum dots in an out-of-plane magnetic field.

Let us set B=0 in Eq.(\ref{eff}),
\begin{eqnarray}
H_{2D}^{(0)} = \varepsilon(k_x) + \frac{m \omega_x^2}{2} (x^2 + 2\beta)  \ .
\label{eff1}
\end{eqnarray}
In semiclassical limit, $n_x \gg 1$, Eq.~(\ref{eff1}) determines $x^2$ as
function of $k_x$ at a given energy $\varepsilon$.
\begin{eqnarray}
\label{xx}
&&x^2=\frac{2}{m\omega_x^2}[\varepsilon-\overline{\varepsilon}(k_x)]\nonumber\\
&&\overline{\varepsilon}(k_x)=\varepsilon(k_x)+m\omega_x^2\beta(k_x) \ .
\end{eqnarray}
Here $\varepsilon$ is the eigenenergy of the state with
quantum number $n_x$.
The $\omega_x^2$-term in (\ref{xx}) can be safely neglected,
so $\overline{\varepsilon}(k_x)\approx \varepsilon(k_x)$.
Hamiltonian (\ref{eff1})
depends quadratically on $x$, therefore, having in mind
the interchange $k_x\to x$, $x\to k_x$, it is easy to find the
semiclassical eigenfunction of (\ref{eff1}) in the momentum representation.
\begin{eqnarray}
\label{wfsc}
\Psi^2(k_x, \varepsilon) = \frac{1}{N(\varepsilon)
\sqrt{\varepsilon - \overline{\varepsilon}(k_x)}}
\end{eqnarray}
The eigenenergy $\varepsilon$ is determined by the Bohr-Sommerfeld 
quantization condition
\begin{equation}
\label{BZ}
4 \cdot \sqrt{\frac{2}{m\omega_x^2}}\int\limits_0^{k_{max}}
\sqrt{\varepsilon-\overline{\varepsilon}(k_x)}dk_x=2\pi n_x \ ,
\end{equation}
where $k_{max}$ is the turning point in the momentum space, 
$\overline{\varepsilon}(k_{max})=\epsilon$.
The normalization coefficient in Eq.~(\ref{wfsc}) is
\begin{equation}
\label{nn}
N(\varepsilon) = \int\limits_0^{k_{ max}} \frac{dk_x}
{\sqrt{\varepsilon - \overline{\varepsilon}(k_x)}} \ .
\end{equation}
This normalization assumes that the momentum in (\ref{wfsc}) is
always positive, $k_x > 0$.

The Zeeman splitting of the oscillator level is given by
usual perturbation theory with wave function (\ref{wfsc}) and
with the B-term in Eq.~\ref{eff} being the perturbation
\begin{eqnarray}
\label{gg}
&&g(\varepsilon) = \int\limits_0^{k_{max}} \Psi^2(k_x, \varepsilon) [g_{zz}(k_x) 
+ g_\alpha(k_x)] \, dk_x \\
&&g_\alpha(k_x)=4 m^2 \omega_x^2 \alpha(k_x) x^2(k_x) \ , \nonumber
\end{eqnarray}
where $x^2(k_x)$ is defined by Eq.~(\ref{xx}).
We know the function $g(\varepsilon)$ from the numerical diagonalization
of 3D Luttinger Hamiltonian, see Eq.~(\ref{gex}). 
Functions $\Psi(k_x, \varepsilon)$, $\alpha(k_x)$ and $x(k_x)$ have been 
already calculated.
In the next paragraph we explain how to invert the integral equation
(\ref{gg}) and to find the function $g_{zz}(k_x)$.

To solve the integral equation (\ref{gg}) we use a well known mathematical
method developed in classical mechanics for the purpose to determine potential
in terms of known period of motion~\cite{LL1}.
Here we briefly describe the method.
We can rewrite Eq.~(\ref{gg}) as
\begin{equation}
f(\varepsilon) = \int\limits_0^{k_{\rm max}} \frac{g_{zz}(k_x) \, dk_x}
{\sqrt{\varepsilon - {\overline \varepsilon}(k_x)}} \ ,
\label{gp}
\end{equation}
where
\begin{equation}
f(\varepsilon) = N(\varepsilon) 
\left( g(\varepsilon) - \int\limits_0^{k_{max}} \Psi^2(k_x, \varepsilon) 
g_\alpha(k_x) \, dk_x \right)
\label{f}
\end{equation}
is known function.
Changing the integration variable we transform ({\ref{gp}) to
\begin{equation}
f(\varepsilon) = \int\limits_0^{\varepsilon} 
\frac{h({\overline\varepsilon})
\, d{\overline\varepsilon}}{\sqrt{\varepsilon - {\overline\varepsilon}}},
\label{h}
\end{equation}
where
\begin{equation}
h({\overline\varepsilon}) = \frac{g_{zz}(k({\overline\varepsilon}))}
{v({\overline\varepsilon})},
\end{equation}
$v({\overline\varepsilon}) = \partial {\overline\varepsilon(k)}/\partial k$ 
is the velocity.
Next we integrate Eq.~(\ref{h}) over $\varepsilon$ with the kernel
$1/\sqrt{\eta-\varepsilon}$, where $\eta$ is an external energy variable.
\begin{equation}
\label{et}
\int\limits_0^{\eta} \frac{f(\varepsilon) \, d\varepsilon}{\sqrt{\eta -\varepsilon}} = 
\int\limits_0^{\eta} \int\limits_0^{\varepsilon} \frac{h({\overline\varepsilon}) \, 
d{\overline\varepsilon} d\varepsilon}{\sqrt{\eta - \varepsilon} 
\sqrt{\varepsilon - {\overline\varepsilon}}} = \pi \int\limits_0^\eta h(\varepsilon) \, d\varepsilon  \ .
\end{equation}
Finally, differentiating Eq.~(\ref{et}) over $\eta$, we find the function
$g_{zz}(k)$:
\begin{equation}
g_{zz}(k) = \frac{v(\eta)}{2\pi\sqrt{\eta}} 
\int\limits_0^1  [f(\eta y) + 2 \eta y f'(\eta y)]\frac{dy}{\sqrt{1-y}} \ ,
\label{gperp}
\end{equation}
where $\eta$ and $k$ are related as $\eta={\overline\varepsilon}(k)$.

Eq.~(\ref{gperp}) solves the inverse problem. So, having the result (\ref{gex})
of the numerical diagonalization of 3D Luttinger Hamiltonian and using 
Eq.~(\ref{gperp}) we find the out-of-plane g-factor.
Plots of $g_{zz}(k)$ for parabolic and rectangular quantum wells in
GaAs and InAs are presented in Fig.~\ref{gzzGaAs}.
\begin{figure}[h]
\includegraphics[width=0.22\textwidth]{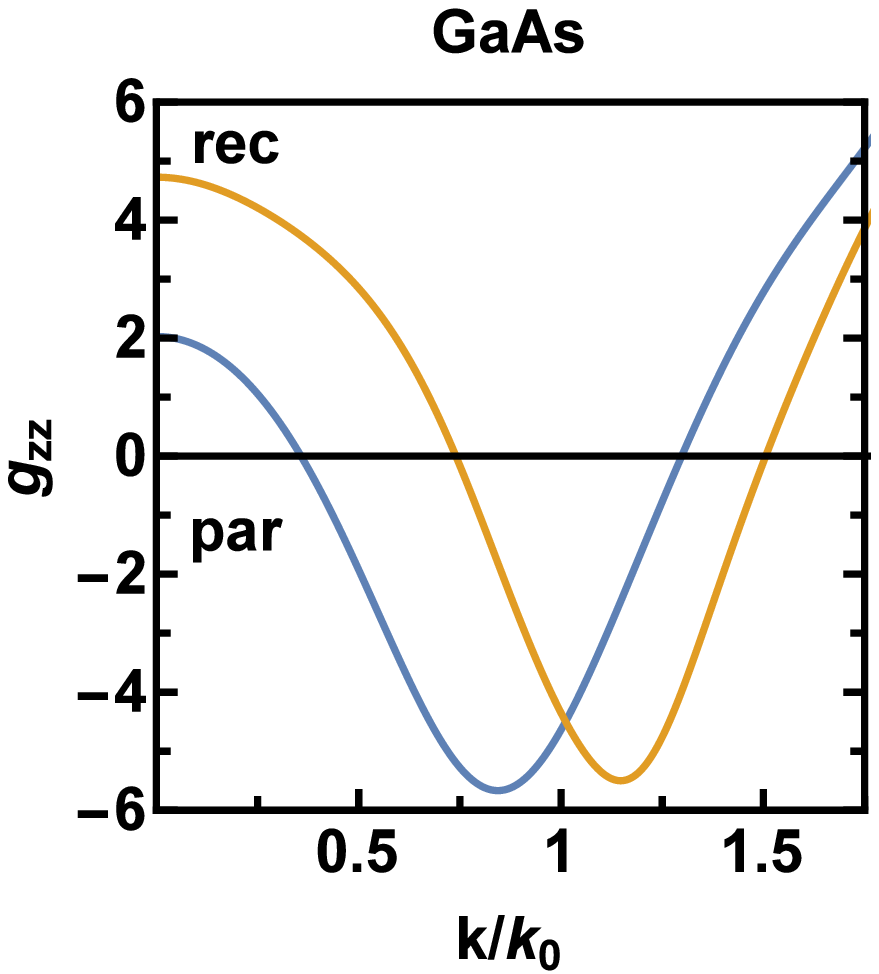}
\includegraphics[width=0.22\textwidth]{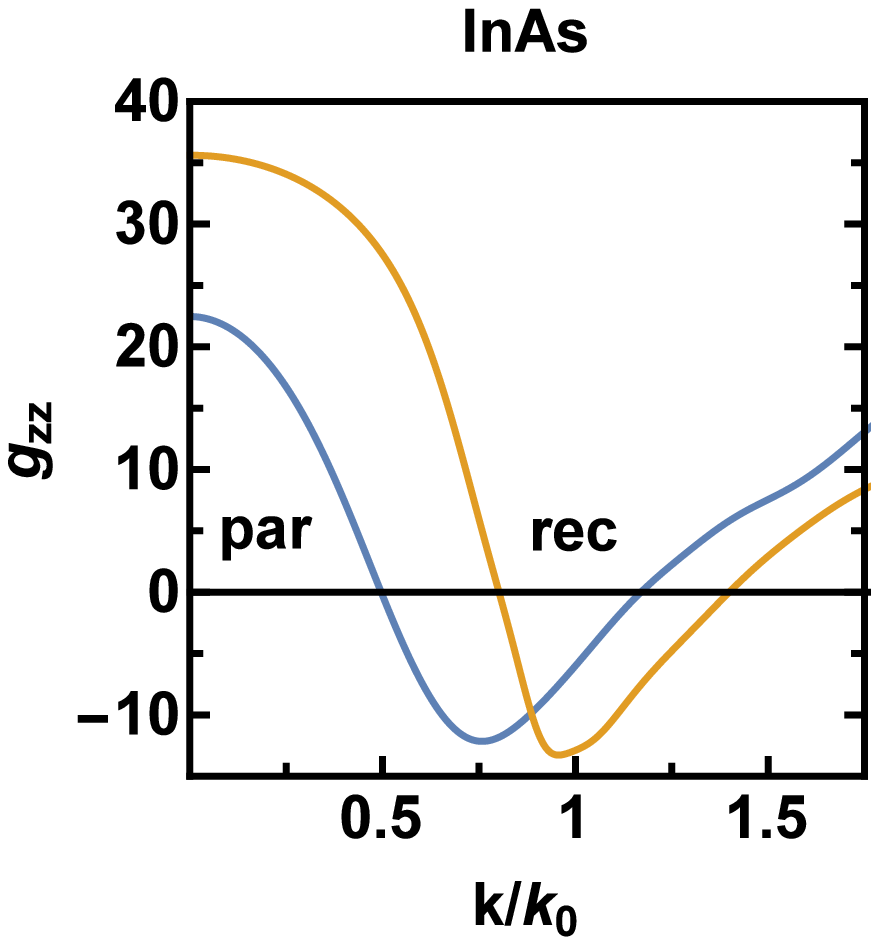}
\caption{Out-of-plane g-factor $g_{zz}$ as function of the in-plane momentum.
The g-factor is presented for
parabolic and rectangular quantum wells in GaAs and InAs.
}
	\label{gzzGaAs}
\end{figure}

Typical experimental densities correspond to the in-plane momentum $k = 0.5 \div 1.5$. According to Fig.~\ref{gzzGaAs}, this region contains two points where $g_{zz}$ changes its sign and these points can be achieved experimentally. Absolute value $|g_{zz}| \approx 5$ at $ k \approx 1$ is consistent with experimental data, Ref.~\cite{ashwin1}

\section{conclusions} \label{conc}
We perform the 3D $\to $ 2D dimensional reduction of the Luttinger 
Hamiltonian for holes in zinc-blende semiconductors
in the presence of a symmetric quantum well, a smooth lateral
gate potential, and an uniform external magnetic field.
Our results are applicable to all kinds of two-dimensional symmetric
semiconductor heterostructures. The effective 2D Hamiltonian, 
Eq.~(\ref{heff}), is written
as a Ginzburg-Landau-type gradient expansion in gradients of the
lateral potential.  We develop general methods and techniques to calculate
parameters of the effective Hamiltonian as functions of the hole momentum.
We specifically present numerical results for the parabolic quantum well
and for the infinite rectangular quantum well in GaAs and InAs.
In the paper we obtain the following results.
(i) We develop the method of calculation and calculate g-factors
for the in-plane direction of the magnetic field.
In particular, we point out that there are two kinematically 
different g-factors, $g_1$ and $g_2$. An important consequence of two 
different g-factors is anisotropic magnetic response in presence of 
anisotropic lateral gate
potential. The g-factors as functions of the hole momentum are plotted
in Fig.~\ref{nu}.
(ii) We develop the method of calculation and calculate the spin-orbit 
interaction and the Darwin interaction related to the lateral gate potential.
The functions $\alpha$ (spin-orbit) and $\beta$ (Darwin) are
plotted in Fig.~\ref{SOI} versus the hole momentum.
(iii) We develop the method of calculation and calculate the $g_{zz}$-factor
for the out-of-plane direction of the magnetic field.
We also point out that in presence of a gate potential (quantum
point contact or a quantum dot) magnetic response is not only due to
$g_{zz}$, there is a part of the response related to the gate
potential which is also calculated. 
The plot of $g_{zz}$ versus momentum is presented in
Fig.~\ref{gzzGaAs}.

\section{acknowledgments}
We thank A.~Hamilton and A.~Srinivasan for useful discussions and interest to the work. We are also grateful to L.~E.~Golub and U.~Z${\rm \ddot{u}}$elicke for valuable communications. The work has been supported by the Australian Research Council grant DP160103630.

 \end{document}